\documentstyle[12pt]{article}
\begin{document}

\title{{\bf Strong Phase Correlations of Solitons of Nonlinear Schr\"odinger
Equation}}
\author{A. G. Litvak, V. A. Mironov, and A. P. Protogenov \\ %EndAName
{\em Institute of Applied Physics, Russian Academy of Sciences}\\ {\em 46
Ulyanov Street, 603000 Nizhny Novgorod, Russia}}
\date{}
\maketitle

\begin{abstract}
We discuss the possibility to suppress the collapse in the nonlinear 2+1 D
Schr\"odinger equation by using the gauge theory of strong phase
correlations. It is shown that invariance relative to $q$-deformed
Hopf algebra with deformation parameter $q$ being the fourth root of unity
makes
the values of the Chern-Simons term coefficient, $k=2$, and of the coupling
constant, $g=1/2$, fixed; no collapsing solutions are present at those
values.
\end{abstract}

\baselineskip=18pt

\section*{1. Introduction}

The characteristic feature of the nonlinear Schr\"odinger equation
\begin{equation}
\label{1}i\partial _t\nabla\Psi =-\frac 12\Delta \Psi -g|\Psi |^{2\alpha }\Psi
\end{equation}
in the case of the positive coupling constant, $g$, and a sufficiently high
level of nonlinearity , $\alpha \geq 2/d$, determined by spatial
dimensionality $d$ of the problem, is existence of collapsing solution \cite
{Tala66,Litv91}. At those self-focusing solutions \cite{Tala66} the
Hamiltonian
\begin{equation}
\label{2}H=\int\limits_{\cal M} d^{2}x\left( \frac 12|\Psi |^2-\frac
{g}{\alpha+1}
|\Psi |^{2(\alpha+1) }\right)
\end{equation}
is unbounded from below (for $d=2$ and for a manifold $\cal M$ being
the whole plane).
This fact hinders using such a solution for
calculating the partition function for the ensemble of nonlinear modes under
consideration. In terms of the situation of general position there is no
prohibition for existence of collapsing solution in model (1). At the same
time , it is difficult to see the aforesaid phenomenon as an example that
leads us beyond the framework of general statistical principles in the case
of multi-soliton field configurations. That means that within the theory of
the field in plasma and in nonlinear optics, in which the nonlinear
Schr\"odinger equation is actively used \cite{Zakh72,Litv91,Litv86},
there is a possibility to stabilize the collapse.

So far it has stayed absolutely unclear, what is the way to solve the
problem associated with unboundedness of the Hamiltonian at collapsing
solutions in the three-dimensional space.
In $2+1 D$ systems the situation is
somewhat different. Peculiarity of $2+1 D$ systems is connected primarily with
realization of fractional statistics in them \cite{Marh76,Wilc82}.
The reason for realization of fractional statistics is purely
topological. It is caused totally by the fact that the $2D$ configuration
space, $M$, is multi-connected, due to which the fundamental group
$\pi_{1}(M)$ coincides with the braid group $B$.
 From the $2+1 D$ point of view,
the complicated topological pattern manifests
itself in the phenomenon of brading of world lines distributed over
the plane of degrees of freedom. Generally speaking, correlations of
that kind are so strong that properties of one realization
of field distribution depend on all the rest configurations in the whole
system.

Within the long-wave approximation the way, which is often used
for account of such strong correlations, requires taking
into account the Chern-Simons term in the Lagrangian
and replacing
usual derivatives with co-variant ones that contain gauge
potential. Applying that approach to the nonlinear Schr\"odinger
equation in the $2+1 D$ case yielded unexpected effects.
It turned out that the solution found by Jackiw and Pi \cite{Jack90}
contains considerable contribution from the gauge field
(see also Ref. \cite{Kim90}). It is thought generally \cite{Wilc89}
that the Chern-Simons interaction modifies statistics, while
making no influence on dynamics properties of the system.
Further we will give arguments showing that those two points of
view \cite{Jack90,Wilc89} do not contradict each other in reality.

One of the major results of the work of Jackiw and Pi \cite{Jack90}
was the conclusion about existence of such a value
for the coupling constant, $g=1/k$, for which the dangerous contribution
$-|\Psi |^{4}$ for $d=2$ forming the collapsing solution is reduced.
We would like to pay attention to this phenomenon for the fixed
value of the coefficient before the Chern-Simons term, $k=2$.
It is corresponded by the maximum value of the coupling constant,
$g=1/2$, in this state of strong phase correlations. By this,
we turn to a structure of deformed Hopf's algebra, on which the
Chern-Simons interaction is based, that is, the so-called
``quantum group'' with the deformation parameter being a root
of unity.

In terms of the most complete account of the symmetry
realized in the ground state, the mentioned value of
the coefficient, $k=2$, makes the statistic and dynamic approaches
self-consistent, as well as emphasized symmetry invariance of the model in
the sense of deformed Hopf's algebra.

Physical realization of the possibility of collapse suppression for classic
wave fields is determined by a number of restrictions that
will be discussed further.
We do not separate now the classic and quantum cases and consider
them from the common point of view. By that we mean that field distributions
in the quantum region (after corresponding re-calculation of scales) are
transferred into the classic region without variations in the qualitative
picture. The reason for that being valid lies, in our opinion, in the
topologic nature of strong phase correlations of $2+1 D$ wave fields
considered in the infrared limit.

The qualitative pattern of the phenomena under consideration in the
$2+1 D$ case will be valid, as we think, for any complex field.
Specifically, interesting results for the $SU(2)\times U(1)$ model for a
similar approach, but from a somewhat different point of view have been
recently obtained in Ref. \cite{Bimo94}. Details of
relativistic soliton configurations, as well as influence of
boundary conditions were the object of recent studies
in Refs. \cite{JaKW90,Bara94}, respectively.

\section*{2. Soliton structure}

To discuss the effect of collapse stabilization and symmetry properties of
the main state we will briefly review basic expressions of
Ref. \cite{Jack90}.
The gauged $2+1D$ nonlinear Schr\"odinger equation is the equation
for motion of the system with the following Lagrangean density:
\begin{equation}
\label{3}{\cal L} = - \frac{k}{8}\: \varepsilon^{\mu \nu \lambda} a_{\mu}
f_{\nu \lambda} + i\Psi^{\ast}(\partial_{t}+ia_{0})\Psi -
\frac {1}{2}|{\bf D}\Psi |^{2} + \frac {g}{2}|\Psi |^{4}.
\end{equation}
Here $a_\mu $ is gauge potential that parametrizes the Chern-Simons term
$U(1)$, $\hbar =e=m=c=1$, ${\bf D}\Psi =\left( {\bf \nabla}
- i{\bf a}\right) \Psi $ is
covariant gradient; the Lagrangian density of the inherent gauge field
$f^{2}_{\alpha \beta}$ is omitted for the sake of simplicity.
Account of that term  does not change the qualitative results
\cite{Bara94}.

The Hamiltonian for system (\ref{3})
\begin{equation}
\label{4}H = \frac {1}{2}\int\limits_{\cal M} d^{2}x \left (|{\bf
D}\Psi|^{2}-g|\Psi|^{4} \right)
\end{equation}
with the use of Bogomolny's decomposition \cite{Bogo76} have
the following form:
\begin{equation}
\label{5}H = \frac {1}{2}\int\limits_{\cal M} d^{2}x
\left (|\left (D_{1}-iD_{2}\right )\Psi |^{2} - (g-k^{-1})|\Psi|^{4} \right).
\end{equation}
Term $\frac 12{\bf \nabla }\times {\bf J}$ with current density
${\bf J}=(1/2i)\left (\Psi ^{\ast}{\bf D}\Psi -
\Psi{\bf D}\Psi^{\ast} \right)$ , which was omitted
when going from Eq.~(\ref{4}) to Eq.~(\ref{5}), yields zero contribution to
Eq.~(\ref{5}) after integration.

We see that at $g=1/k$ the Hamiltonian is determined positively. Moreover,
it equals zero at the solutions of the self-duality equation,
\begin{equation}
\label{6}D_1\Psi = iD_2\Psi.
\end{equation}

Solutions of the self-duality equation, (\ref{6}), have the
form \cite{Jack90} of
\begin{equation}
\label{7}\Psi \left( r,\vartheta \right) = \frac {2n{\sqrt k}}{r}
\left (\left (\frac {r}{r_{0}}\right)^{n} +
\left (\frac {r_{0}}{r}\right)^{n} \right)^{-1}e^{i(1-n)\vartheta},
\end{equation}
\begin{equation}
\label{8}a_i\left( r,\vartheta \right) =
\frac {2n\epsilon_{ij}x_{j}}{r^{2}}
\left (1 + \left (\frac {r_{0}}{r}\right)^{2n} \right)^{-1}.
\end{equation}
In Eqs.~(\ref{7}) and (\ref{8}) $r_0$ is arbitrary scale, integer $n\geq 2$
determines flux $\Phi = \int d^{2}x \; f_{1\, 2} = 4\pi n$ of
the statistic gauge field and whole number of solutions.

The condition of $n\geq 2$ follows from the requirement that the conformal
generator should be finite \cite{Jack90}.
Using the conformal transformation,
$t^{-1}\rightarrow t^{-1}+a$, ${\bf r}\rightarrow {\bf r}/(1+at)$ and
static solutions (\ref{7}), one can obtain time-dependent distribution
\begin{equation}
\label{9}\Psi \left( t,{\bf r} \right) = (1 - at)^{-1}
exp [-iar^{2}/2(1-at)] \Psi \left({\bf r}/(1-at) \right).
\end{equation}

Multi-soliton solution of Eqs.~(\ref{6}) contains \cite{Jack90}
$4n$ parameters of the
parameters that set the phase, scale, and two spatial coordinates for each
of the $n$ solitons. For all those field configurations at $g=1/k$ the
energy equals zero. (Note that the Chern-Simons term in (\ref{3}) contains
first derivatives and yields, therefore, no contribution to the energy).
That means that $4n$ parameters together with the value of coefficient $k$
set dimensionality of the space of degeneration of the ground state, and,
thus, the number of zeroth modes.

Calculation of the partition function,
\begin{equation}
Z=\sum\limits_{all\; states}e^{-\beta H},
%\label{10}Z=\dsum\Sb \text{over} \\ \text{all states}\endSb e^{-\beta H},
\end{equation}
which is equal in this case to the statistic weight,
is reduced to calculation
of the contribution, which arises from the zeroth modes, to the determinant
of the operator that describes small Gaussian deviations
from solutions (\ref
{7}) and (\ref{8}). The problem of calculating the statistic weight has not
been solved yet and lies beyond the scope of the present paper, so we will
concentrate on the problem of collapse stabilization.\thinspace

In terms of calculating the partition function the dynamic details of field
distributions ~(\ref{7}) and (\ref{8}) are not very important. Of essential
significance are only the number of independent parameters and the fact that
the Hamiltonian is bounded from below. Therefore, from the statistical point
of view the problem of the influence of the gauge field on specific
realization of field $\Psi $ becomes less acute. The main role of the gauge
field is transferred to condition $g=k^{-1}$ that limits the Hamiltonian
from below. That condition of compensation in the space of parameters is
supplementary to the self-duality equations, which are usually sufficient for
description of field distributions in the state with zero energy, with the
topologic charge differing from zero, and with finite action. Let us clear
up the reason for compensation of the nonlinear term, $-\left| \Psi \right|
^4$ in Hamiltonian~(\ref{4}) in more detail.

Nonlinear term $-\left| \Psi \right| ^4$ in the Hamiltonian exists as a
quasi-classic limit of the interaction Hamiltonian for the second-quantized
description of Bose particles with the pairwise attracting $\delta $-function
interaction.
It is well known from energy estimations from below \cite{Wegn71,Kada71}
that there is
no prohibition for a collapse in the Bose system. Meanwhile, for fermions
Pauli's principle of exception limits the energy from below due to fermion
repulsion \cite{Kada71} and prevents a collapse.

In $2+1 D$ systems the classification of field distributions is known as
fractional or intermediate statistics and is built in correspondence to
irreducible representations of the braid group
(see, e.g., Ref.\cite{Eina90}). Its
Abelian and, in the general case, non-Abelian irreducible representations are
parametrized by the phase $\exp (2\pi i\Phi )$, determined by flux $\Phi $
of the gauge field. In the non-Abelian case irreducible one-dimensional
representations that belong to a discrete center of the group are
parametrized by flux $\Phi =P/Q$ with mutually prime numbers $P$ and $Q$
\cite{Hats94}. The flux of the gauge field describes
coherent adiabatic rotation
dynamics of Bose particles with hard cores. The condition of the hard
core relfects the effect of {\em distinguishability of loop links}
constructed by the linked world lines projected on a plane. The chiral
character of motion, which arises due to spontaneous breaking
of the angular momentun projection being orthogonal to the plane, leads
to partial Bose repulsion depending on the linking parameter. That effect of
strong phase correlations with the repulsind character inherent for them
prevents a collapse under the condition of compensation, $g=k^{-1}$.

Are there any outstanding values of $k$ in equation $g=k^{-1}$? In other
words, what is the value of $g$ under the conditions of collapse
suppression? To obtain an answer, let us turn to analysis of symmetry
properties of system~(\ref{3}).

Besides a wide choice of spatio-temporal symmetries \cite{Jack90},
the Chern-Simons sector of model~(\ref{3}) contains
an algebraic structure of Hopf's deformed algebra.
In terms of physics the latter describes the laws of
addition of dynamic integrals of motion for fusion of strongly correlated
fluctuations. As the deformation parameter, $q$, of Hopf's
algebra in our case serves the adiabatic phase, $\exp (2\pi i\Phi )$, which
is equal, at $\Phi =P/Q$, to the $Q$th root of unity. Further we will
discuss the connection between numbers $Q$ and $k$, and now we will
emphasize specific character of representations of Hopf's deformed algebra
with the deformation parameter being a root of unity.

For $q$ being a root of unity, Hopf's $q$-deformed algebra contains a rich
center \cite{Roch89} that ``maintains'' its irreducible
representations. By that,
in the general case the central elements are not independent but satisfy
some polynominal relations \cite{Roch89}.
There is also come consequence of $q$%
-deformed algebras obtained from the initial one by division into invariant
relations of central elements.

Generally speaking, there is no reversible universal $R$-matrix of Hopf's
deformed algebra at roots of unity. However, for the above quotients there
are always transpositions of tensor products of representations, i.e., there
are always $R$-matrices on representations. Some of those quotients have
finite dimensionality and a universal $R$-matrix.

In a seminal paper \cite{Arna94} Arnaudon has obtained
the answer to the following question: on which condition
are the different quotients of initial Hopf's algebra
Hopf-equivalent? In the case of them being equivalent, the universal
$\cal R$-matrix of some of them can be transformed into a universal
$\cal R$-matrix of the others.
In Ref. \cite{Arna94} it was proved
that it happens only when $Q=4$. Moreover, Ref.~\cite{Arna94} gave accurate
expressions for the automorphisms and for the
transformed universal $\cal R$-matrices for that case.

When studying the connection between the conformal field theory and
Chern-Simons's topological theory it was stated \cite{Guad90} that
for $SU(N)$ of the
gauge field $Q=k+N$. For $Q=4$ and minimal limit, $N=2$ , of the non-Abelian
symmetry we have $k=2$. In the absence of the effect of shift to the value
of $N=2$ coefficient $k$ would have equaled 4. Note also that at $k=2$ both
the angular momentum, $J$, measured in units of 2$\pi $, and the number of
particles, ${\cal N} = \int |\Psi |^{2}\;d^{2}x = 4{\pi }|k|n$,
coincides with the number 4$n$ of zeroth modes.

An integral value of coefficient $k$ for the non-Abelian Chern-Simons term
does not cause any objections, since it is dictated by invariance of the
additional term in it as related to global gauge transformations. We deal
with an Abelian gauge field, for which there is no complete and
final proof of coefficient $k$ being an integer.
Realization of various possibilities in that respect depends
primarily on the boundary conditions related to global
gauge transformations determined over all the $2+1 D$ manifold, i.e., on
its topological structure. In that situation of a
certain arbitrary rule it is
natural to see an Abelian gauge field as a field
belonging to the center of its non-Abelian $SU(2)$ predecessor,
and use, having applied that reduction, $N=2$.
Singling out from the non-Abelian $SU(N)$ gauge theory its Abelian
discrete center $Z_n$ is given in Ref. \cite{Hats94}.
Some other arguments for integral (even) values of $k$ in the Abelian
case are contained in Refs. \cite{Hoso89}.

Even in the case of $k=2$ it is still unclear, what is the value of $g$,
since seemingly, applying only to Eq.~(\ref{1}), we could always change the
scales so that coefficient $g$ would equal any number. So, let us supplement
the motion equation limited with solutions of the self-duality
equation \cite{Jack90},
\begin{equation}
\label{11}\nabla ^2\ln |\Psi |^2 = -\frac {2}{k}(2kg -1)|\Psi|^{2},
\end{equation}
with the constraint that reflects the Gauss law,
$\varepsilon _{\alpha \beta}f_{\alpha \beta } = k^{-1}|\Psi|^{2}$ , i.e.,
\begin{equation}
\label{12}\nabla ^2\ln |\Psi |^2 = -\frac {2}{k}|\Psi|^{2}.
\end{equation}
We see that at $k=2$ Eqs.~(\ref{11}) and (\ref{12}) are compatible only in
the case of $g=1/2$. The same value of parameters
satisfy the condition of $kg=1$ of collapse prevention.

\section*{3. Conclusion}

When $d=2$, $g=1/2$, and
\begin{equation}
\label{13}2i\partial _t\Psi + \nabla^{2}\Psi + |\Psi|^{2}\Psi = 0,
\end{equation}
Eq.~(\ref{1}) can be seen as a local limit of description of more realistic
situations. One of such examples,
\begin{equation}
\label{14}2i\partial _t\Psi + \nabla^{2}\Psi + {\rho }\Psi = 0,
\end{equation}
\begin{equation}
\label{15}\rho _{tt} - \nabla^{2}\rho + \nabla^{2}|\Psi |^{2} = 0,
\end{equation}
describes sound motion of density $\rho $ in the presence of an external
ponderomotive force \cite{Zakh72}. Within a weak limit
($|\Psi |^2\rightarrow 0$)
distribution of density relaxes towards the quasi-equilibrium value, $\rho
=|\Psi |^2$, and we come back to Eq.~(\ref{13}).

Another situation arises when scale $l$ of variation of density $\rho $ does
not coincide with the scale of the field, $\Psi $. It is corresponded by the
system of equations that includes Eq.~(\ref{14}) and equation
\begin{equation}
\label{16}\rho - l^{2}\nabla^{2}\rho - |\Psi|^{2} = 0.
\end{equation}
At $l\gg 1$ Eqs.~(\ref{14}) and (\ref{16}) describe such wide-scale
distributions of concentration, $\rho $, that scale $l$ houses several
solitary distributions of field $\Psi $.

This situation is corresponded by the integral connection of field $\rho $
with field $\Psi $ in Eq.~(\ref{14}), and hence, by effective weakening of
the nonlinearity degree, $\alpha $. Therefore, at $l\neq 0$ and $d=2$ the
nonlinearity degree $\alpha <2$, and the problem of collapse does not arise.

However, in connection with the possibility of sound generation (see Eqs.~(%
\ref{14}) and (\ref{15})) another difficulty appears. The matter is that in
some Chern-Simons system there is no soft mode. Such a situation
occurs, e.g., in the systems, in which the
Fractional Quantum Hall Effect is realized.
The sound mode is replaced by a quasi-particle with a finite gap in the
long-wave limit, a so-called magnetoroton. Its existence reflects rigidness
of the system of random and, at the same time, coherent vortex
distributions of the field.

In order to retain phase coherence, which may be broken by sound reduction
of the described topologic rigidness, in our case, one has to neglect the
first term in Eq.~(\ref{15}) to reconstruct local connection $\rho =|\Psi |^2
$. It follows from comparison of the terms in Eqs.~(\ref{14}) and (\ref{15})
that it is possible when $r_0\ll 1$ and $|\Psi |^2r_0^2\sim 1$. Those
conditions may be satisfied only in the case of a great number, $n\gg 1$, of
solitons, and, therefore, at a sufficiently large number, ${\cal N}\sim n$,
of particles. For the statistical description the latter condition is assumed
satisfied.

Finally, we would like to make the following comment: as it has been said
already, the universal integer, $n$, numbers both the solitons and their
angular momenta. For field distributions, in Eqs.~(\ref{7}) and (\ref{8})
the angular momentum $n\geq 2$. The monopole value, $n=0$, is excluded,
since this value corresponds to the absence of vortices and consideration of
the plane as a non-punctured manifold. On the other hand, the traditional
collapse \cite{Tala66,Zakh72} is singularity of the solutions
of the nonlinear Schr\"odinger equation in the $n=0$ channel.
It is possible that the case of $n=0$
corresponds to the phase state with vortex singularities compensated due to
disorientation, and the collaps is prevented, e.g., due to saturation of
nonlinearity. However, in the framework of the considered model without
nonlinearity saturation we have to assume that $n\geq 2$.

In conclussion, we used hidden symmetry of the gauged 2+1D
nonlinear Schr\"odinger equation relative to action of $q$-deformed Hopf
algebra with deformation parameter $q=\exp \left[ 2\pi i/(2+k)\right] $,
and gave arguments for the fact that at the minimum possible value,
$k=2$, the value of coupling constant, $g$, equals 1/2, for which the
condition $gk=1$ of collapse suppression is fulfilled.

\section*{Acknowledgments}

We are grateful to A.D. Yunakovsky for valuable discussions of
the problem. This work was supported in part by Grant No. R8K000
from the ISF.
One of the authors (A.P.) would like to thank Professor
Abdus Salam, the International Atomic Energy Agency
and UNESCO for hospitality at the the International Centre
for Theoretical Physics, Trieste.

\end{document}